\documentclass[a4paper,11pt]{article}
\pdfoutput=1 
\usepackage{jheppub} 
\usepackage[normalem]{ulem}
\usepackage{graphicx,subfigure}
\usepackage{float}
\usepackage{dcolumn}
\usepackage{bm}
\usepackage{pdfpages}
\usepackage{epstopdf}
\usepackage{mathrsfs}
\usepackage{amssymb,amsmath,amsfonts,latexsym}
\usepackage{nicefrac}
\usepackage[colorlinks=true,linktocpage=true,linkcolor=blue,citecolor=blue]{hyperref}
\usepackage[utf8]{inputenc}
\usepackage{lipsum}
\usepackage{nicefrac}
\usepackage{slashed}
\usepackage{mathtools}
\usepackage{multirow}
\usepackage{caption}
\usepackage{lineno}
\usepackage{orcidlink}

\long\def\comment#1{ }

\definecolor{kkcolor}{rgb}{1,0,0}

\newcommand{\beq}{\begin{eqnarray}}
\newcommand{\eeq}{\end{eqnarray}}

\newcommand{\be}{\begin{eqnarray*}}
\newcommand{\ee}{\end{eqnarray*}}

\newcommand{\rmd}{{\rm d}}
\newcommand{\dd}{{\rm d}}
\newcommand{\rme}{{\rm e}}

\def\rmR{{\rm Re}}

\def\q{{\rm q}}

\newcommand{\nn}{\nonumber\\ }

\newcommand{\Leff}{L_{\rm eff}}

\def\dd{{\rm d}}

\newcommand{\bem}{\begin{multline}}
\newcommand{\eem}{\end{multline}}
\newcommand{\beg}{\begin{gather}}
\newcommand{\eeg}{\end{gather}}

\newcommand{\ben}{\begin{eqnarray*}}
\newcommand{\een}{\end{eqnarray*}}

\newcommand{\secn}[1]{Section~1}
\newcommand{\appn}[1]{Appendix~1}
\long\def\comment#1{ }

\def\and{\quad\text{and}\quad}

\def\q{{\boldsymbol q}}
\def\0{{\boldsymbol 0}}

\def\lbs{{\boldsymbol l}}
\def\k{{\boldsymbol k}}

\def\n{{\boldsymbol n}}
\def\x{{\boldsymbol x}}

\begin{document}

\title{Transverse momentum broadening of medium-induced cascades in expanding media}

\author[a]{Souvik Priyam Adhya\orcidlink{https://orcid.org/0000-0002-4825-2827}}
\emailAdd{souvik.adhya@ifj.edu.pl}
\author[a,b]{Krzysztof Kutak\orcidlink{https://orcid.org/0000-0003-1924-7372}}
\emailAdd{krzysztof.kutak@ifj.edu.pl}
\author[c]{Wies{\l}aw P{\l}aczek\orcidlink{https://orcid.org/0000-0002-9678-9303}}
\emailAdd{wieslaw.placzek@uj.edu.pl}
\author[d]{Martin Rohrmoser\orcidlink{https://orcid.org/0000-0003-2311-832X}}
\emailAdd{rohrmoser.martin1987@gmail.com}
\author[e]{Konrad Tywoniuk\orcidlink{https://orcid.org/0000-0001-5677-0010}}
\emailAdd{konrad.tywoniuk@uib.no}

\affiliation[a]{Institute of Nuclear Physics, Polish Academy of Sciences,
ul.\ Radzikowskiego~152, 31-342 Krakow, Poland}
\affiliation[b]{ Brookhaven National Laboratory, Physics Department, Bldg. 510A, 20 Pennsylvania Street, 30-059 Upton, NY 11973 USA}
\affiliation[c]{Institute of Applied Computer Science, Jagiellonian University, ul.\ {\L}ojasiewicza~11, 30-348 Krakow, Poland}
\affiliation[d]{Faculty of Material Engineering and Physics, Cracow University of Technology, ul.\ Podchor\c{a}\.zych~1, 30-084 Krakow, Poland}
\affiliation[e]{Department of Physics and Technology, University of Bergen, 5007 Bergen, Norway}
\date{\today}
\preprint{IFJPAN-IV-2022-19}

\abstract{
In this work, we explore the features of gluonic cascades in static and Bjorken expanding media by numerically solving the full BDIM evolution equations in longitudinal momentum fraction $x$ and transverse momentum $\k$ using the Monte Carlo event generator {\sf MINCAS}. Confirming the scaling of the energy spectra at low-$x$, discovered in earlier works, we use this insight to compare the amount of broadening in static and expanding media. We compare angular distributions for the in-cone radiation for different medium profiles with the effective scaling laws and conclude that the out-of-cone energy loss proceeds via the radiative break-up of hard fragments, followed by an angular broadening of soft fragments. While the dilution of the medium due to expansion significantly affects the broadening of the leading fragments, we provide evidence that in the low-$x$ regime, which is responsible for most of the gluon multiplicity in the cascade, the angular distributions are very similar when comparing different medium profiles at an equivalent, effective in-medium path length. This is mainly due to the fact that in this regime, the broadening is dominated by multiple splittings. Finally, we discuss the impact of our results on the phenomenological description of the out-of-cone radiation and jet quenching. 
}
\maketitle
\flushbottom

\section{Introduction}

In ultra-relativistic collisions of heavy ions, nuclear matter is subjected to such extreme conditions that it leads to the creation of a quark-gluon plasma (QGP) \cite{Busza:2018rrf}. This state of matter is opaque to jet propagation, a phenomenon referred to as ``jet quenching'' \cite{dEnterria:2009xfs,Mehtar-Tani:2013pia,Blaizot:2015lma}. This phenomenon was early proposed as a sensitive observable to extract medium properties in Refs.~\cite{Gyulassy:1990ye,Wang:1991xy}, and was observed in collider experiments at RHIC~\cite{Adler:2002tq} and LHC~\cite{Aad:2010bu}. 
Jet quenching has been addressed using methods like AdS/CFT~\cite{Liu:2006ug,Ghiglieri:2020dpq}, 
semi-analytical calculations  \cite{Baier:1994bd,Baier:1996vi,Zakharov:1996fv,Zakharov:1997uu,Zakharov:1999zk,Baier:2000mf,Gyulassy:2000fs,Guo:2000nz,Barata:2020rdn,Barata:2020sav,Mehtar-Tani:2019ygg,Mehtar-Tani:2019tvy}, kinetic theory \cite{Baier:2000sb,Jeon:2003gi,Arnold:2002ja,Ghiglieri:2015ala,Kurkela:2018wud} and, finally, Monte Carlo and other numerical methods \cite{Salgado:2003gb,Zapp:2008gi,Armesto:2009fj,Schenke:2009gb,Lokhtin:2011qq,Casalderrey-Solana:2014bpa,Kutak:2018dim,Blanco:2020uzy,Caron-Huot:2010qjx,Feal:2018sml,Andres:2020vxs}.

In recent  years a lot of studies were focused on the out-of-cone radiation or equivalently transverse-momentum dependence of produced mini-jets \cite{Blaizot:2013hx}.  On a theoretical level, this problem leads to the generalization of  the BDMPS-Z  framework\cite{Baier:2000mf,Baier:2000sb,Jeon:2003gi,Zakharov:1996fv,Zakharov:1997uu,Zakharov:1999zk,Baier:1994bd,Baier:1996vi,Arnold:2002ja,Ghiglieri:2015ala}, and to the formulation of the BDIM equations for gluons \cite{Blaizot:2012fh} and combined system of quarks and gluons \cite{Blanco:2020uzy}.  
One of the actively investigated problems is to  account for the expansion of the medium 
together with turbulent dynamics of the jet losing energy.
In particular, in Ref.~\cite{Adhya:2019qse}, the medium-modified gluon splitting rates for different profiles of the expanding partonic medium, namely the profiles for the static, exponential, and Bjorken expanding medium were calculated.
This generalized the rate equation for the energy distribution that accounts for multiple soft scatterings with a time-dependent transport coefficient characterizing the expanding medium. This allowed quantifying the sensitivity of the inclusive jet suppression on the way how the medium expands.  In the subsequent study \cite{Adhya:2021kws}, the framework was generalized to account for quarks. Apart from that, the initial-state nuclear effects, vacuum-like emissions as well as coherence effects were taken into account. This allowed to achieve a reasonable description of the nuclear modification of jets.

In this paper, we would like to  generalize the framework of Ref.~\cite{Adhya:2019qse, Adhya:2021kws} by accounting for the transverse-momentum dependence of mini-jets produced during the jet quenching. This will amount to generalizing the BDIM equations to the expanding medium-dependent case. The result will allow us to determine the effect of expansion on transverse-momentum spectra of fragmenting jets and to investigate the interplay of dynamics  of expansion and re-scatterings. One of the objectives of this work is to study the impact of different effects of the medium expansion on the radiation and broadening revealed by the scaling analysis in the effective evolution parameters. Secondly, we study the radial distributions of gluons in bins of $x$ (or energy) to show where the hard and soft emissions dominate. This reveals the medium kinematical scales where the splitting and the broadening dominate, respectively, as well as differences between various medium profiles. 

In a very recent paper \cite{Mehtar-Tani:2022zwf}, the authors studied similar physical quantities as in this work through a kinetic theory approach with thermalization effects included for an infinite static medium. In this paper, we present results for a more realistic description using a time-dependent splitting rate that captures the interplay of the formation time of the radiation and the medium length for the static media. Next, we highlight qualitative differences between the media profiles by including the Bjorken expanding media and  explore the sensitivity of the physical quantities on the time for the onset of the jet quenching.

The paper is organized as follows. In Sec.~\ref{sec:setup}, we revisit the single-gluon distributions and collinear splitting kernels for the expanding as well as static medium. We also introduce the in-medium gluon evolution equation.
Next, in Sec.~\ref{sec:spec_long}, we discuss the scaling features of a purely radiative cascade for different medium profiles, present formal solutions, and compare them to numerical evaluations in a dedicated Monte Carlo event generator {\sf MINCAS}. The established scaling allows corroborating the concept of an equivalent \textit{effective} in-medium path length $\Leff$, that leads to the same low-$x$ spectrum in different medium scenarios. In Sec.~\ref{sec:spec_full}, we study fully differential spectra of mini jets both in the static medium as well as in the medium undergoing the Bjorken expansion, obtained via numerical evaluations in {\sf MINCAS}. In particular, we analyze the dependence of the fragmentation function on a polar angle. From the full distributions in the longitudinal momentum fraction $x$ and the transverse momentum $k_T$, alternatively $x$ and the polar angle $\theta$, we discuss three key quantities: a) the average $k_T$ distribution; b) the $x$-distribution within a certain cone limited by $\theta < \theta_{\rm max}$; and c) the angular distribution in specific bins of $x$. Finally, in Sec.~\ref{sec:gluons-in-cone}, we perform an in-depth study of the fraction of gluons in a given $x$-range that remain within a cone. In Sec.~\ref{sec:conclusions}, we summarize our work and explore the possible qualitative impact on jet quenching phenomenology. 

\section{Medium-induced cascades in expanding media}
\label{sec:setup}

In this work, we study two classes of medium profiles: a static, non-expanding medium and a medium following the Bjorken expansion.

\paragraph{Static medium:} In this case, we consider the temperature of the medium to be independent of time, $T = T_0$.

\paragraph{Bjorken expanding medium:} Assuming one-dimensional longitudinal expansion, we use the following form for the temperature evolution as a function of the proper time $t$,
\beq
\label{eq:temp_evol}
T(t) = \begin{cases} 0 & {\rm for } \quad t<t_0 \,, \\ T_0 \left(\frac{t_0}{t}\right)^{\frac{1}{3}} & {\rm for} \quad t_0 \le t \le L+t_0 \,, \\ 0 & {\rm for} \quad t > L+t_0  \,,\end{cases}
\eeq
where the reference time $t_0$ is a free parameter. 

In addition, we allow for two different initial time values $t_0$ in the Bjorken scenario, resulting in a total of three studied cases.
Further discussion about the expanding media can be found in Refs.~\cite{Adhya:2019qse,Salgado:2002cd, Adhya:2021kws, Andres:2019eus, Salgado:2003gb}.

We consider the gluon distribution
\begin{equation} 
D(x,\k,t) \equiv x \frac{\rmd N}{\rmd x \rmd^2 \k } \,,
\end{equation}
where $x$ is the longitudinal momentum fraction and $\k=(k_x,k_y)$ is the transverse momentum. The evolution equation for this distribution in a dense medium, neglecting quark contributions, reads \cite{Blaizot:2014rla}
\begin{align}
\label{eq:BDIM2}
\frac{\partial}{\partial t} D(x,\k,t) &= \frac{1}{t_\ast}
\int_0^1 \rmd z\, {\cal \tilde K}(z, t) \left[\frac{1}{z^2}\sqrt{\frac{z}{x}}\, D\left(\frac{x}{z},\frac{\k}{z},t\right)\theta(z-x) - \frac{z}{\sqrt{x}}\, D(x,\k,t) \right] \nn
&+ \int \frac{\rmd^2 \lbs}{(2\pi)^2}\,C(\lbs,t)\, D(x,\k-\lbs,t)\,.
\end{align}
The above evolution equation describes the interplay between the collinear splittings (first two terms on the r.h.s.\ of the equation) and diffusion in momentum space (the last term). In this equation, we consider the case when the transverse-momentum transfer and broadening during the branching is neglected. The collinear branching kernel ${\cal \tilde K}(z,t)$ will be discussed in detail below. The transverse-momentum dependence comes, nevertheless, from both the elastic scattering, as described by the elastic collision kernel $C(\lbs,t)$, and multiple splittings, see the terms in the first line of Eq.~\eqref{eq:BDIM2}. Finally, the characteristic time
\begin{equation}
    t_\ast \equiv \frac{1}{\bar \alpha} \sqrt{\frac{p_0^+}{\hat q_0}} \,,
\end{equation}
with $\bar \alpha = \alpha_s N_c/\pi$ and $p_0^+$ being the (light-cone) energy of the initial jet particle, determines the stopping time of the jet at the initial, or \textsl{reference}, value of the jet quenching parameter $\hat q_0$, which will be specified in more detail below. In this work, we have not included the thermalization effects for $x \le 0.01$ as in Ref.~\cite{Mehtar-Tani:2022zwf}.

The elastic collision kernel $C(\lbs,t)$ is given by
\begin{equation}
\label{eq:Cq}
C(\lbs,t) = w(\lbs,t) - \delta^{(2)}(\lbs) \int \rmd^2\lbs'\, w(\lbs',t)\,,
\end{equation}
where $w(\lbs,t) \propto n(t) \rmd^2 \sigma_\text{el}/\rmd^2\lbs$ is proportional to the density of the medium $n(t)$ and the in-medium elastic cross section $\rmd^2 \sigma_\text{el}/\rmd^2\lbs$. When the medium expands, the density drops reducing the impact of elastic momentum transfer. Although it is less important, screening effects, contained in the elastic cross section, will also typically be hampered by the expansion\footnote{To be precise, we consider the \textsl{uniform} expansion, for extensions see e.g.\ Ref.~\cite{Barata:2022krd}.}.

Considering purely gluon systems, we use the so-called  Hard Thermal Loop (HTL) approximation for $w(\lbs,t)$ at the leading order:
\begin{equation}
    w(\lbs,t) = \frac{N_c g^2 m_D^2 T}{\lbs^2(\lbs^2+m_D^2)} = \frac{4\pi \, \hat q}{\lbs^2 (\lbs^2 + m_D^2)} \,,
\label{eq:wltHTL}
\end{equation}
where the medium is characterized by the temperature $T$ and the Debye mass $m_D$, and we have not written explicitly their time dependence. We have also introduced the jet transport coefficient, which is given by $\hat q = \alpha_s N_c m_D^2 T$ with $g^2 = 4\pi \alpha_s$. The Debye mass in the QCD medium at the leading order is given by
\beq
m_D^2 = g^2T^2\left(\frac{N_c}{3}+\frac{N_f}{6}\right) = \frac32 g^2 T^2 \,.
\eeq

In addition to the above perturbative collision kernels, one can also study the effect of using scattering potentials from non-perturbative extraction from the lattice QCD \cite{Schlichting:2021idr} as well as the NLO HTL contributions \cite{Caron-Huot:2010qjx}. 
In a static medium, $m_D^2 = m_{D0}^2 \equiv 3/2 g^2T_0^2 $ and $\hat q = \hat q_0 \equiv \alpha_s N_c m_{D0}^2 T_0$ are simply constant, whereas in the Bjorken model these parameters, together with the temperature, are time-dependent and given by
\begin{equation}
    \hat q(t) = \hat q_0 \left(\frac{T(t)}{T_0} \right)^3 \,, \qquad m_D^2(t) = m_{D0}^2 \left(\frac{T(t)}{T_0} \right)^{2} \,,
\end{equation}
where the time evolution of the temperature is given in Eq.~\eqref{eq:temp_evol}.

Next, we turn to the splitting kernel. It is related to the in-medium splitting rate as $\mathcal{\tilde K}(z, t) = t_\ast\, \rmd I/(\rmd z\, \rmd t)$. This rate has recently been evaluated numerically in a static medium in Ref.~\cite{Andres:2020vxs}. It has also been investigated in  analytical resummation schemes that cover the whole kinematical phase space \cite{Mehtar-Tani:2019tvy,Mehtar-Tani:2019ygg,Barata:2020sav,Andres:2020vxs,Isaksen:2022pkj}. We shall  use  it together with the so-called ``harmonic oscillator'' approximation which is valid for soft emissions in a large medium \cite{Isaksen:2022pkj}. The main features of soft, multiple emissions are captured by this approximation. It is however worth pointing out that the scattering kernel, discussed in the preceding paragraphs, includes hard scatterings as well. A systematic study of improving the splitting kernel to include such rare interactions, which affects the early-time behavior of the rate, is left for the future.

For a time-dependent splitting kernel, a pertinent question is whether the relevant time should refer to the time between subsequent splittings or whether it should be counted from the beginning of the cascade. The former scenario is significantly more complicated, since it would involve a complicated interplay between subsequent emissions, and remains an open challenge (see  Ref.~\cite{Barata:2021byj} for an exploratory study). This problem applies both to the static and time-dependent media, see below. In Eq.~\eqref{eq:BDIM2}, subsequent splittings are assumed to be independent and the \textsl{global} time is always counted from the beginning of the cascade up to its end.

In the static medium, it is easily identified from the medium-induced spectrum
\cite{Baier:1996kr, Baier:1996sk, Zakharov:1996fv, Zakharov:1997uu, Adhya:2019qse, Adhya:2021kws}, and reads
\begin{align}
\label{eq:rate_static}
\mathcal{\tilde K}^\text{static}(z,t) &= {\cal K}(z) \, \rmR \left[(i-1) \tan \left(\frac{1-i}{2} \kappa(z) \tau \right) \right] \,,\nn
&= \mathcal{K}(z)\, 
\frac{\sinh(\kappa(z) \tau) - \sin(\kappa(z) \tau)}{\cosh(\kappa(z) \tau) + \cos(\kappa(z) \tau)}\,,
\end{align}
where $\tau \equiv t/t_\ast$, $\kappa(z) = \sqrt{[1-z(1-z)]/[z(1-z)]}$ and the time-independent part is
\begin{equation}
    {\cal K}(z) = \frac{\kappa(z) P_{gg}(z)}{2 N_c} \,,
\end{equation}
where the (un-regularized) Altarelli-Parisi splitting function reads $P_{gg}(z) = 2N_c\, [1-z(1-z)]^2/[z(1-z)]$.

In the Bjorken expanding medium, the spectrum is given by Refs.~\cite{Baier:1998yf,Salgado:2003gb,Arnold:2008iy}, resulting in the rate \cite{Adhya:2019qse, Adhya:2021kws}
\begin{align}
\label{eq:rate_bjorken}
\mathcal{\tilde K}^\text{Bjorken}(z,\tau,\tau_0) =  \mathcal{ K}(z)\,\sqrt{\frac{\tau_0}{\tau}}\,
\text{Re} \left[ (1-i) \frac{J_1(z_L) Y_1(z_0) - J_1(z_0) Y_1(z_L) }{J_1(z_0) Y_0(z_L) - J_0(z_L) Y_1(z_0)} \right].  
\end{align}
where $\tau_0 \equiv t_0/t_\ast$, $J_{\alpha}(\cdot)$ and $Y_{\alpha}(\cdot)$ are the Bessel functions of the first and second kind, respectively, and
\begin{align}
z_0 &= (1-i) \kappa(z) \tau_0 \,,\\
z_L &= (1-i) \kappa(z) \sqrt{\tau_0 \tau}\,.
\end{align}
At late times, the factor $\rmR\big[\ldots \big]$ behaves similarly to the static case, saturating at 1. In this case, the main effect of the expansion comes from the factor $\sqrt{\tau_0/\tau}$ which arises directly from considering the static rate with the time-dependent $\hat q$. Note that, in this case, the final time $\tau$ corresponds to $\tau \equiv (t_0+L)/t_\ast$, where $L$ refers to the path length in the medium. It is also interesting to note the additional dependence of the rate on $t_0$ in the Bjorken case. A study of the role of the $t_0$ has recently been done in Ref.~\cite{Adhya:2021kws} on the rapidity ratio and $v_2$ of the jets. In the subsequent sections, we shall further explore the possible physical implications of $t_0$ regarding the scaling laws.

We consider a plasma with the initial temperature $T_0 = 0.4\,$GeV ($N_c=N_f=3$) and the initial momentum to be $p_0^+ = 100\,$GeV. The coupling constant is fixed by $\bar \alpha=0.3$, which corresponds to $g \approx 1.9869$. For the above input parameters we get $m_{D0} = 0.97\,$GeV, $\hat{q}_0 = 1.81\,$GeV$^2$/fm and, finally, $t_\ast = 11.0\,$fm. 
In addition, to facilitate a numerical solution, we impose the minimal value on the $x$ variable at $x_{\rm min} = 10^{-4}$.

Since we present the results in terms of $k_T\equiv |\k|$ dependence, we introduce the distribution
\begin{equation}
    \Tilde{D}(x,k_T,t) = \int_0^{2\pi} 
    \rmd \phi_{\k}\, k_T D(x,\k,t) \,,
    \label{eq:Dtilde}
\end{equation}
such that
\begin{equation}
    D(x,t) = \int_0^{\infty} \rmd k_T\, \Tilde{D}(x,k_T,t) \,,
\end{equation}
corresponds to the final longitudinal momentum, or in short energy, spectrum.

We solve the full evolution equation \eqref{eq:BDIM2} for a collinear kernel using the initial condition of the evolution of the single-particle $D(x,\k,t=0) = \delta (1-x)\delta(\k)$. 
For the results presented in the rest of the paper, we use the numerical solutions of the evolution equation that takes into account the transverse-momentum broadening as well as the splittings in an effective way as already demonstrated in Refs.~\cite{Kutak:2018dim,Blanco:2020uzy,Blanco:2021usa} for the static, infinite media.
To this end, we apply the Markov Chain Monte Carlo (MCMC) algorithm implemented in the event
generator {\sf MINCAS}, as described in Ref.~\cite{Kutak:2018dim}, with the necessary extensions.
These extensions account for the time dependence of the splitting kernels 
given in Eqs.~\eqref{eq:rate_static} and \eqref{eq:rate_bjorken} as well as of the collision kernel 
according to Eq.~\eqref{eq:wltHTL}. Such a dedicated MCMC algorithm proved to be efficient in solving
the evolution equation \eqref{eq:BDIM2}, even for the complicated Bjorken-expanding medium case.

\section{Scaling in the energy spectrum}
\label{sec:spec_long}

Now, integrating Eq.~\eqref{eq:BDIM2} over the transverse momentum $\k$, we obtain the evolution equation 
for the energy distribution $D(x,t) = \int \rmd^2 \k \, D(x,\k,t) = x\, \dd N/\dd x$ 
of the gluon cascade \cite{Blaizot:2013hx,Blaizot:2013vha}:
\begin{align}
\label{eq:RateEquation-generic}
\frac{\partial D(x, t)}{\partial t} = \frac{1}{t_\ast} \int_0^1 \dd z \,\mathcal{\tilde K}(z,t) \left[\sqrt{\frac{z}{x}} D\left(\frac{x}{z},t \right) \Theta(z-x) - \frac{z}{\sqrt{x}} D(x,t) \right] ,
\end{align}
in which the collision term has been integrated out to zero. In this work, we attempt to use the static in-medium kernel rates defined in Eqs.~\eqref{eq:rate_static} and \eqref{eq:rate_bjorken} for which the above equation can only be solved numerically. This was previously also analyzed in Ref.~\cite{Caucal:2020uic}, albeit neglecting the finite-size effects in the rate, as contained in Eq.~\eqref{eq:rate_bjorken}. Inclusion of the finite-size effects in the rate was done in Ref.~\cite{Adhya:2019qse} for the pure-gluon cascade, followed by the multipartonic cascades in Ref.~\cite{Adhya:2021kws}. Here, we use the dedicated MCMC algorithm implemented in the {\sf MINCAS} program \cite{Kutak:2018dim} to solve the equations numerically.  

Before we turn to the numerical evaluations, let us first re-visit the scaling laws in the evolution variable among different medium profiles.
The quenching parameter  ${\hat q}(t)$ for expanding medium is time-dependent. The single-gluon emission spectrum for different medium-expansion scenarios possesses scaling features for an average transport coefficient \cite{Salgado:2002cd,Salgado:2003gb}. However, as demonstrated in Ref.~\cite{Adhya:2019qse}, the average scaling in $\hat q$ is valid only in the hard part of the single-gluon spectrum and not relevant for the soft part of the spectrum which contributes the most to the multiplicity, and therefore quenching \cite{Baier:2001yt}, of emitted gluons. Instead, to establish the optimal scaling in the soft sector of the spectra, an ``effective" quenching parameter has been identified \cite{Adhya:2019qse}. In terms of the \textsl{effective} in-medium evolution time $\Leff$, this translates to
\begin{equation}
\Leff = \int_0^\infty \rmd t' \, \sqrt{\frac{\hat q(t)}{\hat q_0}} \,, 
\end{equation}
where the temperature profile is given by Eq.~\eqref{eq:temp_evol}. This establishes a relation between the length traversed in a static medium (on the left-hand side of the equation) and in an expanding medium (on the right-hand side),  which corresponds to the equivalent amount of quenching. For the Bjorken model considered here, the relation reads
\begin{align}
\label{eq:leff}
    \Leff &= 2 \sqrt{t_0} \left( \sqrt{L + t_0} - \sqrt{t_0}\right) \,.
\end{align}
If we assume $L \gg t_0$, we can write $\Leff \approx 2 \sqrt{t_0 L}\, [1+ \mathcal{O}(\sqrt{t_0/L})]$. Since we are interested in the regime of multiple soft scatterings, we consider only the effective scaling.

\begin{table}[t!]
\centering
\begin{tabular}{lccc}
\hline
Medium & $t_0$ [fm] & $L_{\rm eff}$ [fm] & $L$ [fm] \\ \hline
Static & 0.0 & 4.0; 6.0 & 4.0; 6.0 \\ 
Bjorken (early) & 0.6& 4.0; 6.0 & 10.7; 21.0 \\
Bjorken (late) & 1.0& 4.0; 6.0 & 8.0; 15.0 \\ \hline
\end{tabular}
\caption{Values of the initial time $t_0$, the effective length $L_{\rm eff}$, and the actual length $L$ of the evolution for different types of the medium used in Monte Carlo simulations.}
\label{tab:mediumtauxscale}
\end{table}

\begin{figure}[t!]
\centering 
\includegraphics[scale=0.5]{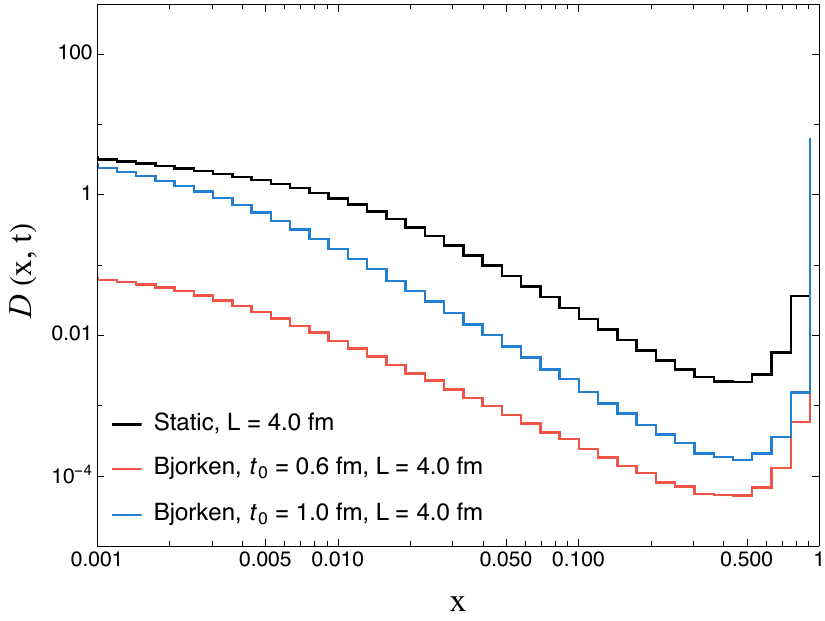}~
\includegraphics[scale=0.5]{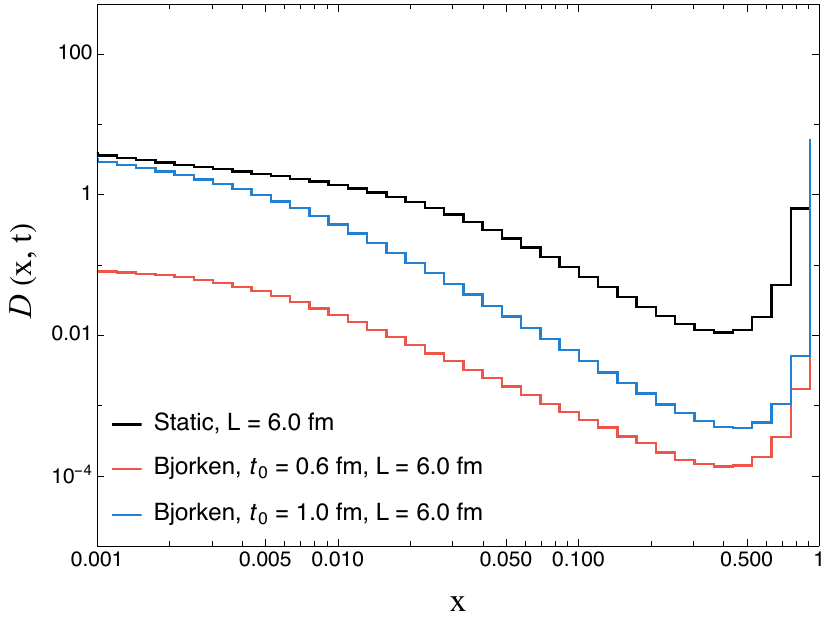}\vspace{3mm}\\
\includegraphics[scale=0.5]{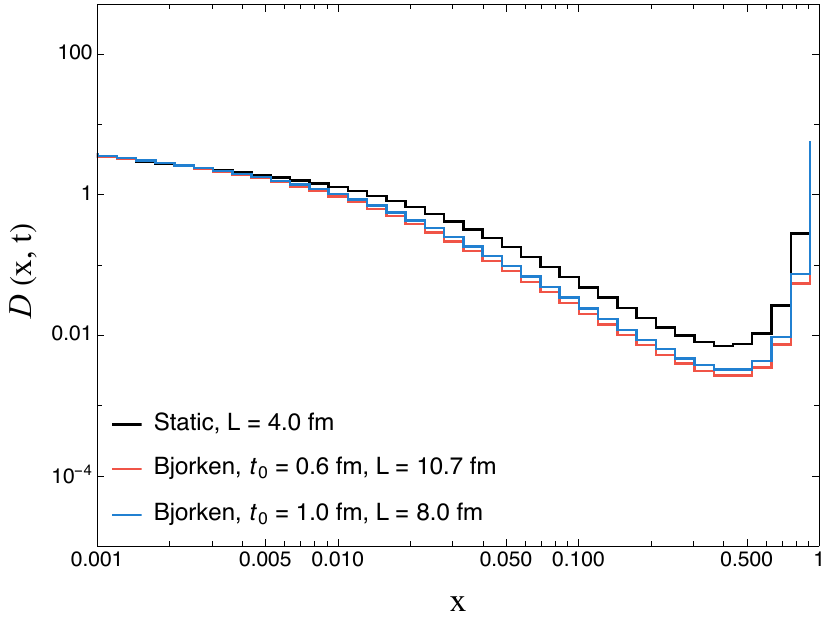}~
\includegraphics[scale=0.5]{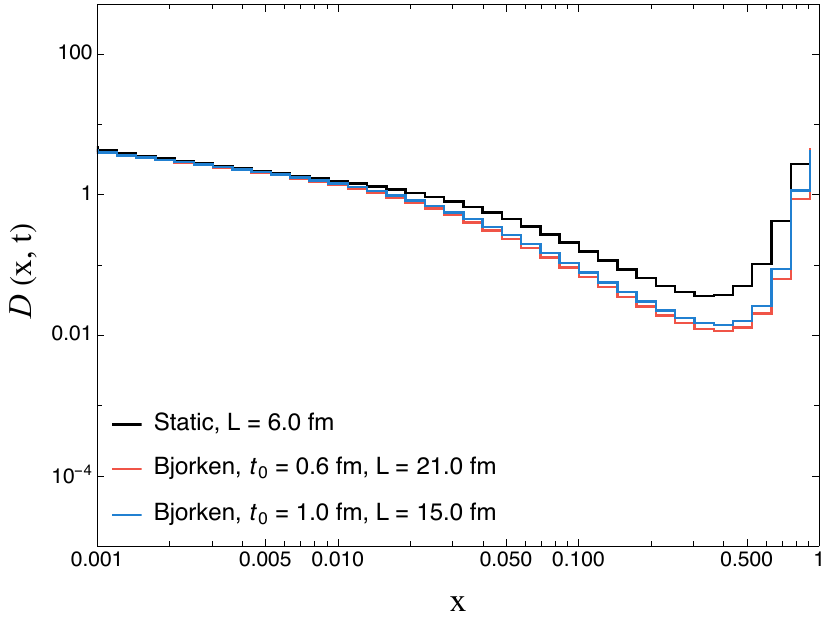}
\caption{The comparison of the gluon energy spectra for different medium profiles. The upper row shows the spectra evaluated at the medium lengths $L = 4\,$fm (left) and $6\,$fm (right), respectively. The lower panels show the scaling of the gluon spectra for the scaling variable $L_{\rm eff}$ corresponding to the two above values of $L$ (see Table~\ref{tab:mediumtauxscale}).} 
\label{fig:spectraDxt}
\end{figure}

The different profiles have been implemented in the Monte Carlo code {\sf MINCAS}, and the results are shown in Fig.~\ref{fig:spectraDxt} for the evolution parameters given in Table~\ref{tab:mediumtauxscale}. In the upper row, we plot the resulting spectra for a fixed length of the medium $L = 4\,$fm and $6\,$fm, with two choices for the Bjorken initial time $t_0 = 0.6\,$fm and $1\,$fm. Naturally, one obtains a much more pronounced evolution for the static medium as compared to the expanding-medium cases. However, we see the appearance of the turbulent-like cascade $\sim 1/\sqrt{x}$ for all the medium profiles. In the lower row of Fig.~\ref{fig:spectraDxt}, the spectra have instead been generated at the equivalent \textit{effective} medium length $L_{\rm eff}$; see the third column in Table~\ref{tab:mediumtauxscale}. One can observe the scaling occurring in the low-$x$ part of the spectra for different medium profiles, thus re-confirming the findings in Ref.~\cite{Adhya:2019qse}.  These scaling features can also be found analytically within a simplified evolution equation, see e.g.\ Ref.~\cite{Caucal:2020uic}.

\section{Fully differential spectra}
\label{sec:spec_full}

We now move to the full solutions of the evolution equation in the energy fraction $x$ and the transverse momentum $k_T\equiv |\k|$. Actually, the full solution of Eq.~\eqref{eq:BDIM2} is 3-dimensional in the variables $(x,\k)$, but its dependence on the azimuthal angle in the transverse-momentum plane $\phi_{\k}$ is trivial, so it is integrated out. In order to facilitate comparisons between the different medium profiles, we shall henceforth consistently plot the distributions at their equivalent \textit{effective} medium lengths $L_{\rm eff}$, see Table~\ref{tab:mediumtauxscale}. The reasoning behind this particular choice for presenting our numerical results will become clear in a moment.

\begin{figure}[t!] 
\centering     
\includegraphics[width=70mm]{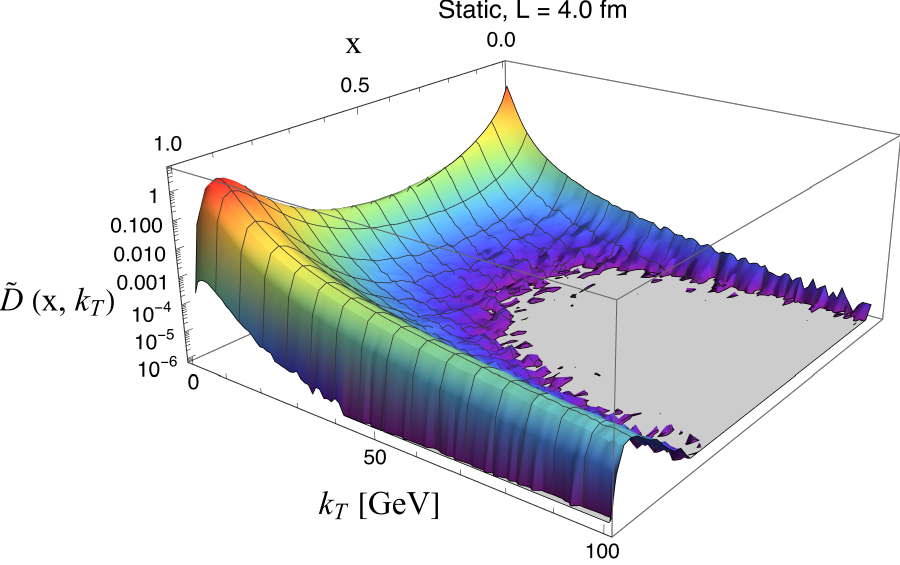}~~~~\includegraphics[width=70mm]{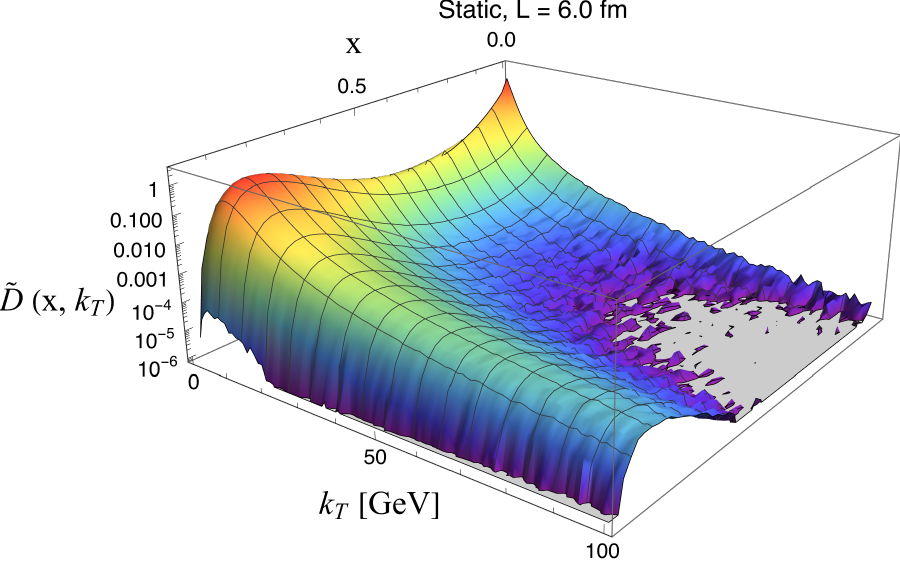}\\
\includegraphics[width=70mm]{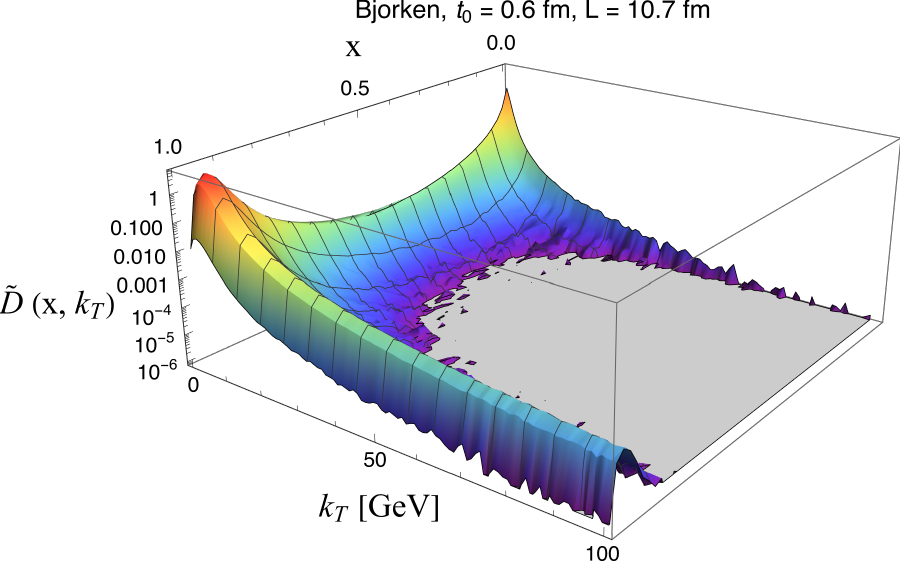}~~~~\includegraphics[width=70mm]{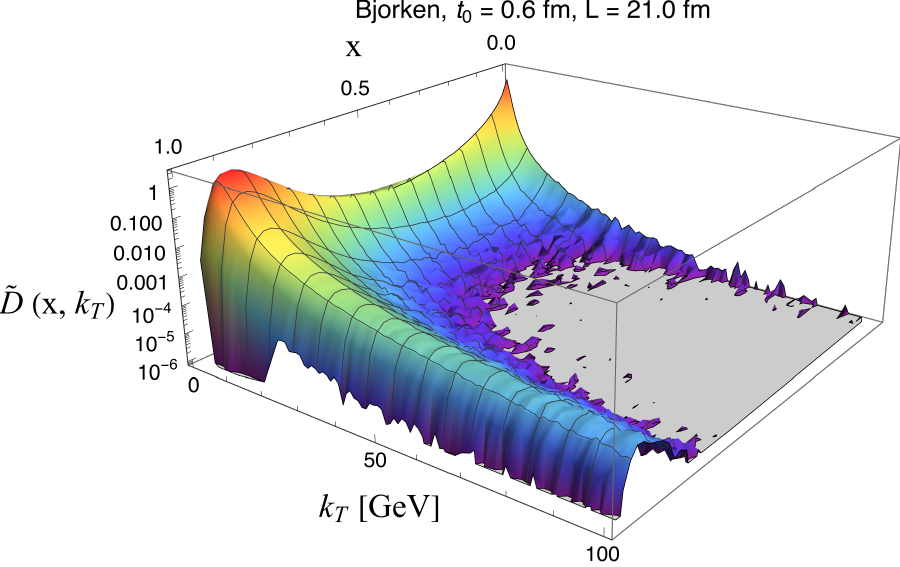}\\
\includegraphics[width=70mm]{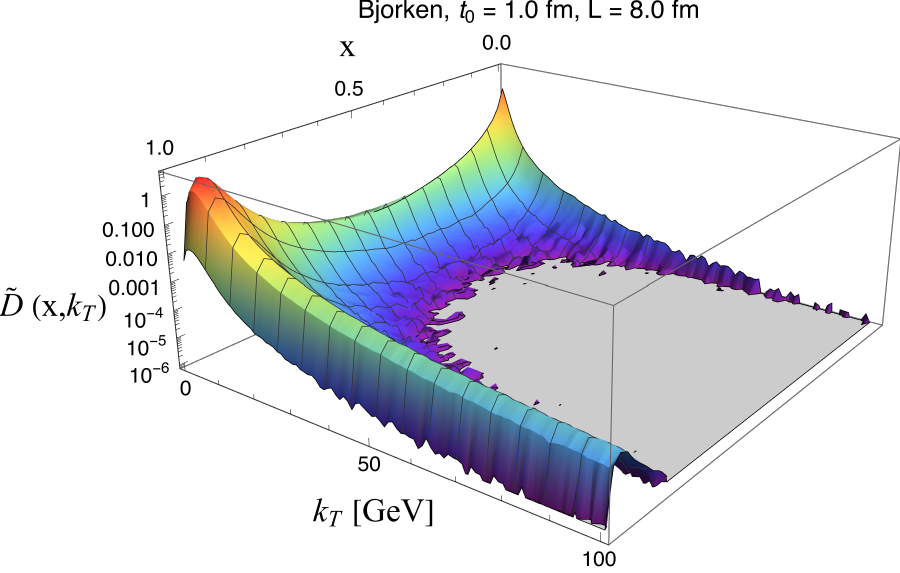}~~~~\includegraphics[width=70mm]{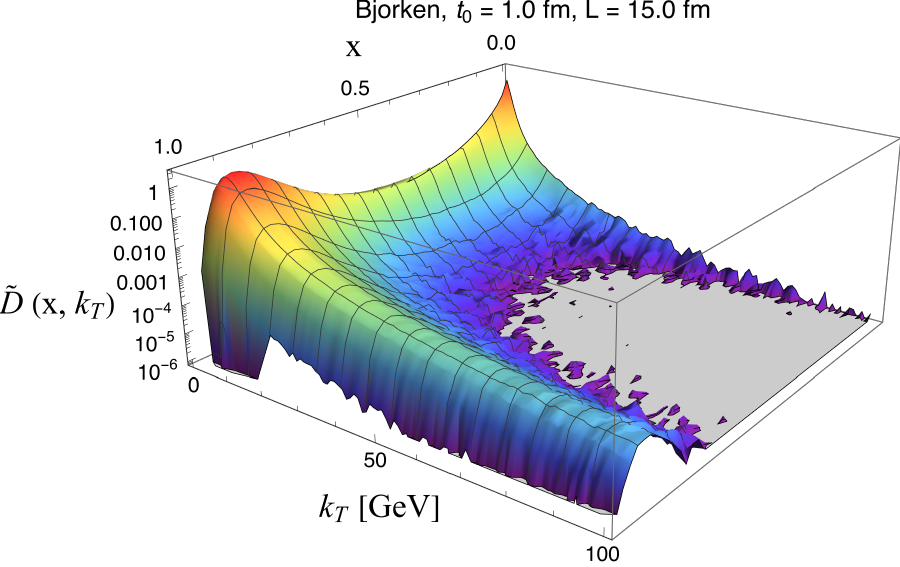}
\caption{The $\tilde{D}(x, k_T)$ distributions for the effective length $L_{\rm eff} = 4\,$fm (left column) and $6\,$fm (right column), see Table~\ref{tab:mediumtauxscale}.}
\label{fig:DxkT3D}
\end{figure}
The 2D differential distributions $\Tilde{D}(x,k_T,t)$ are shown in Fig.~\ref{fig:DxkT3D}. Let us first address the interplay of the hard and soft sectors of jet quenching physics by focusing separately on the low- and high-$x$ regimes of the distributions:
\begin{itemize}
    \item The high-$x$ ($x \sim 1$) regime is dominated by the behavior of the leading fragment in the cascade. The transverse-momentum distribution in this regime is therefore expected to be dominated by multiple soft-gluon scatterings at small $k_T$, leading to the Gaussian profile. At high-$k_T$, this becomes a power-law suppression due to rare hard medium interactions.
    \item In the low-$x$ ($x\ll 1)$ regime, on the other hand, the $k_T$ distribution is narrower and approximately Gaussian. No distinct transition to a power-law behavior is observed for the range of plotted $k_T$ values.
\end{itemize}
These features are qualitatively in agreement with the discussion in Ref.~\cite{Blaizot:2014ula} for the static media. There, the fully differential spectrum was postulated to factorize as $D(x,\k,t) \simeq D(x,t) P(\k,t)$. Let us, therefore, parallel the qualitative observations made directly from the obtained numerical results in Fig.~\ref{fig:DxkT3D} with a discussion based on these analytical estimates. 

At $x\sim 1$, we have $D(x,t) \simeq \delta(1-x)$, while $P(\k,t)$ is simply the well-known broadening distribution for a single particle \cite{Gyulassy:2002yv,Qiu:2003pm,DEramo:2010wup,Barata:2020rdn}, see also Refs.~\cite{Liou:2013qya,Blaizot:2013vha,Caucal:2021lgf,Ghiglieri:2022gyv} for a discussion of radiative corrections. This distribution is defined
\begin{equation}
    P(\k,t) = \int \rmd^2 \x \, \rme^{-i \k \cdot \x}\, \rme^{- \int_0^t \rmd t'\, v(\x,t')} \,,
\end{equation}
where $v(\x,t) = N_c n(t)\int \rmd^2 \q\, (1-\rme^{- i\q \cdot \x}) \sigma(\q)/(2\pi)^2 $ is the so-called dipole cross section, $\sigma(\q) \equiv \rmd \sigma^{\rm el}/\rmd^2 \q$ being the medium elastic scattering potential. This distribution is characterized by two distinct regimes, see e.g.\ Ref.~\cite{Barata:2020rdn}. At small $\k$, we can approximate $v(\x,t) \simeq \hat q(t) \x^2/4$, leading to 
\begin{equation}
    P(\k,t) \simeq \frac{4\pi}{Q^2_s(t)}\,\rme^{- \frac{\k^2}{Q_s^2(t)}} \,,
\end{equation}
where $Q_s^2(t) = \int_0^t \rmd t' \, \hat q(t')$ is the characteristic scale, sometimes called the saturation scale. In a static medium, $Q_s^2 = \hat q_0 L$. This regime is therefore dominated by multiple soft scatterings, leading to the Gaussian broadening. At large $\k$, on the other hand,
\begin{equation}
    P(\k,t) \simeq N_c \sigma(\k) \int_0^t \rmd t'\, n(t') \sim (4\pi)^2 \frac{\alpha_s^2 N_c \int_0^t \rmd t'\, n(t')}{\k^4} \,,
\end{equation}
follows from the generic behavior of $t$-channel exchange. As already indicated in the formulas, this regime is easily generalized to expanding media via the time-dependent quenching parameter $\hat q(t)$ (or the density $n(t)$). Also, it becomes clear that the broadening of the leading fragments is significantly less effective when the medium is becoming dilute compared to when it remains at constant density.

At low-$x$, the situation becomes more complicated. The energy distribution can approximately be described by a turbulent cascade \cite{Blaizot:2013hx}.
The broadening is, however, not easily disentangled in this case due to the structure of the evolution equation \eqref{eq:BDIM2}. For a static medium and in the low-$\k$ regime, it was found \cite{Blaizot:2014ula} that $P(\k,t)$ again becomes Gaussian, however with a width given by $\langle \k^2 \rangle \sim \sqrt{x E \hat q}$. This particular behavior comes about since the $k_T$ distribution in this regime is driven mainly by multiple parton \textit{splittings}. Note also that the $k_T$ distribution is much narrower than in the large-$x$ regime, in line with the results in Fig.~\ref{fig:DxkT3D}. In angular variables, where in the soft limit we can approximate $\theta \sim k_T/(x p_0^+)$, the width is enhanced in the small-$x$ regime.

In an expanding medium, we have not been able to obtain an analytically closed formula in these approximations. However, assuming that the broadening continues to be dominated by multiple \textit{splittings}, we expect to see a similar amount of broadening in the soft sector for the static and expanding medium profiles when evaluated at the equivalent effective evolution time $L_{\rm eff}$. In the next steps, see also Sec.~\ref{sec:gluons-in-cone}, we will study the behavior in the numerical data from {\sf MINCAS} and reveal how wide the jet becomes in expanding media compared to a static one.

\begin{figure}[!ht]
\centering     
\includegraphics[width=70mm]{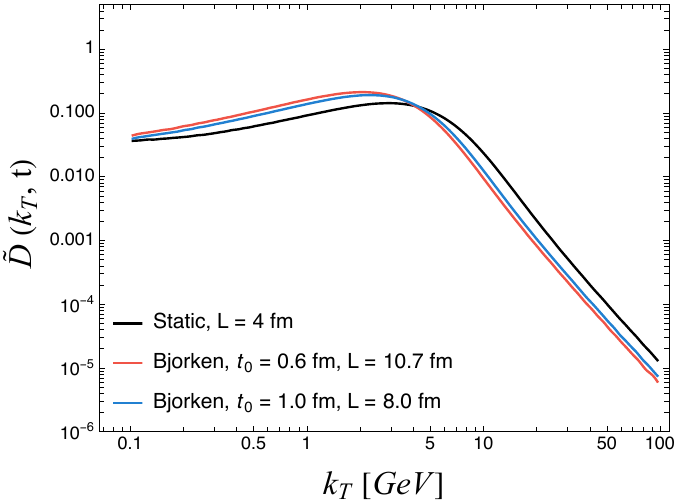}
\includegraphics[width=70mm]{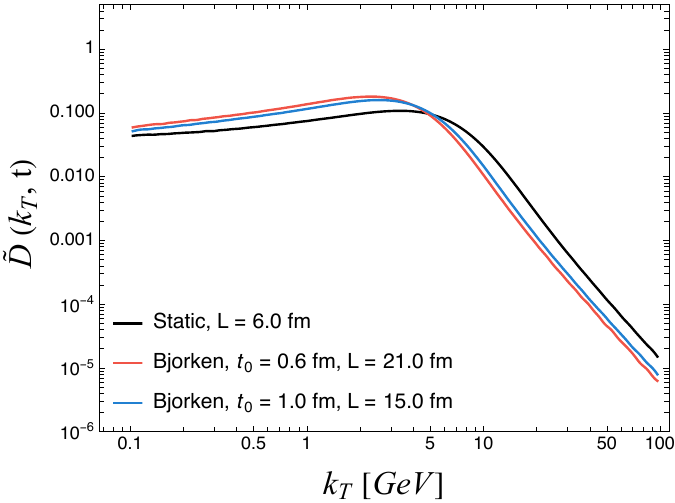}
\caption{The transverse-momentum distribution $\tilde{D}(k_T,t)$ for the effective medium length $L_{\rm eff} = 4\,$fm (left panel) and  $6\,$fm (right panel). 
}
\label{fig:DtildekTallmeds}
\end{figure}

We start with presenting the transverse-momentum spectra of the gluons as obtained from the evolution equation \eqref{eq:BDIM2} in the case of the static medium and the Bjorken expansion. First, we consider the $k_T$ distribution defined as\footnote{In practice, the lower integral limit is given by the internal {\sf MINCAS} parameter $x_{\rm min}$, which for the presented results $=10^{-4}$.} $\tilde D(k_T,t) = \int_0^1 \rmd x \, \tilde D(x,k_T,t)$. This corresponds roughly to the $k_T$ distribution of the average $\langle x \rangle$ fragments in the cascade, and is plotted in Fig.~\ref{fig:DtildekTallmeds} for different media. Due to the Jacobian, the distribution is suppressed at small $k_T$. At intermediate $k_T$ we recognize the characteristic Gaussian peak, followed by a strong, power-law suppression. We have plotted the distributions for the equivalent effective in-medium path lengths, see Table~\ref{tab:mediumtauxscale}. Since $\Leff$ was calculated with the scaling of the soft sector in mind, these distributions, which are mostly sensitive to the large-$x$ fragments, do not scale. Rather, we observe that the interactions occurring in the expanding medium are significantly less efficient in generating large-$k_T$ fragments than in the equivalent static medium.

As mentioned above, the simultaneous process of multiple splittings and scatterings, as encoded in Eq.~\eqref{eq:BDIM2}, sets up a dynamical picture where one has to look for leading and sub-leading contributions of the radiative and collisional processes. At this point, it is useful to switch variables from the transverse momentum $k_T$ to the polar angle $\theta$. This is achieved by the following transformation:
\begin{equation}
     \bar{D}(x,\theta,t) =  xp_0^+\, \tilde D(x,x p_0^+ \theta,t)\,,
\label{eq:Dbar}
\end{equation}
where we have used the small-angle approximation $k_T = x p_0^+\theta$ with $\theta$ being the polar angle\footnote{The general definition of the polar angle in light-cone coordinates is given by $\bar{\theta}(x,\k) =  \arccos\big[(2xp_0^+)^2 - \k^2\big]/\big[(2xp_0^+)^2 + \k^2\big]$, which leads to the distribution $\bar{D}(x,\theta,t) = \int \rmd^2\k\, D(x,\k,t)\,\delta\left(\theta - \bar{\theta}(x,\k)\right)$. We have explicitly checked in our numerical results that the small-angle approximation works well in the angular region considered here, i.e.\ $\theta \leq 0.6$.}. It then naturally follows that the fully inclusive angular distribution is given by $\bar{D}(\theta,t) = \int_0^1 \rmd x \, \bar{D}(x,\theta,t)$. In order to explore the scales where different effects are at play, we separate the problem by looking at the $x$ distribution in specific bins of the angle $\theta$ and vice versa.

\begin{figure}[!ht]
\centering     
\includegraphics[width=70mm]{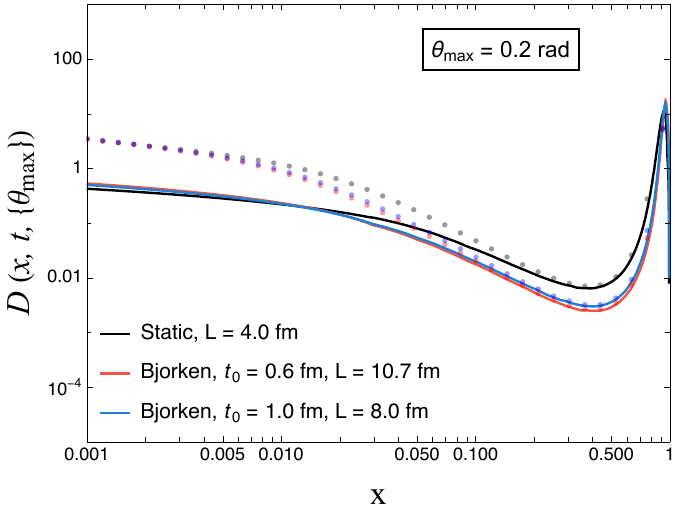}~~~~\includegraphics[width=70mm]{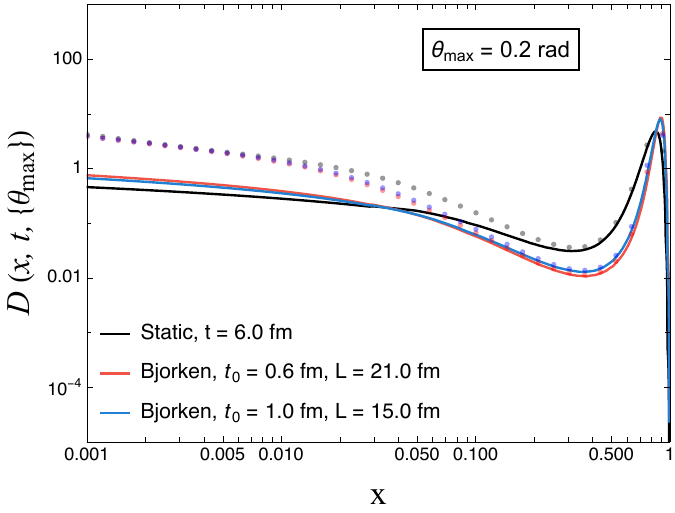}\\
\includegraphics[width=70mm]
{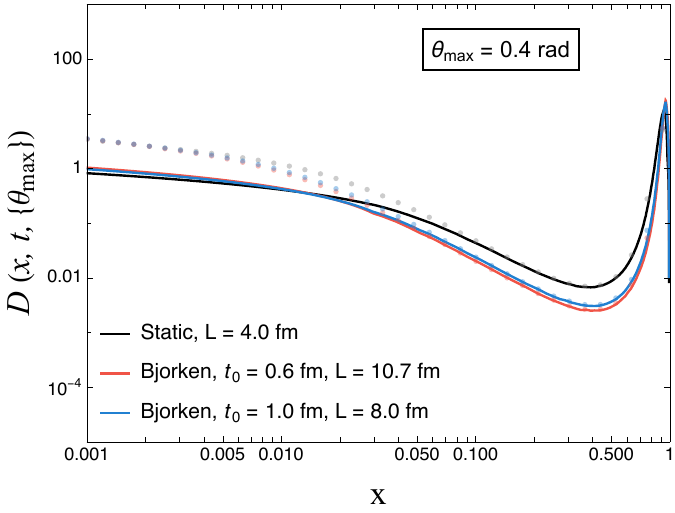}~~~~\includegraphics[width=70mm]{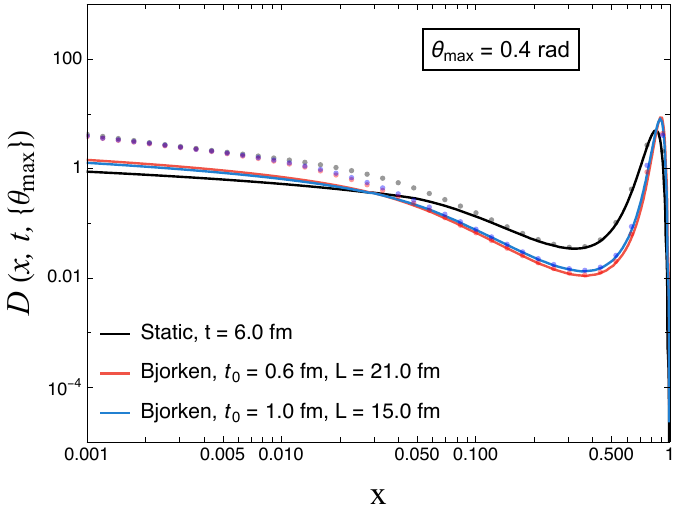}\\
\includegraphics[width=70mm]
{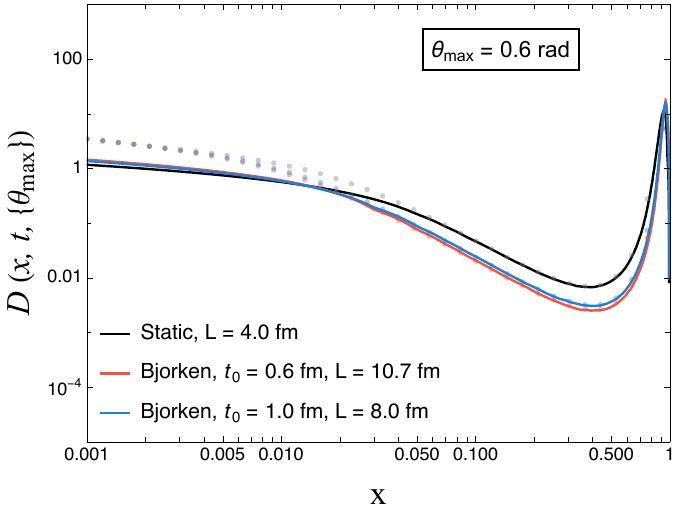}~~~~\includegraphics[width=70mm]{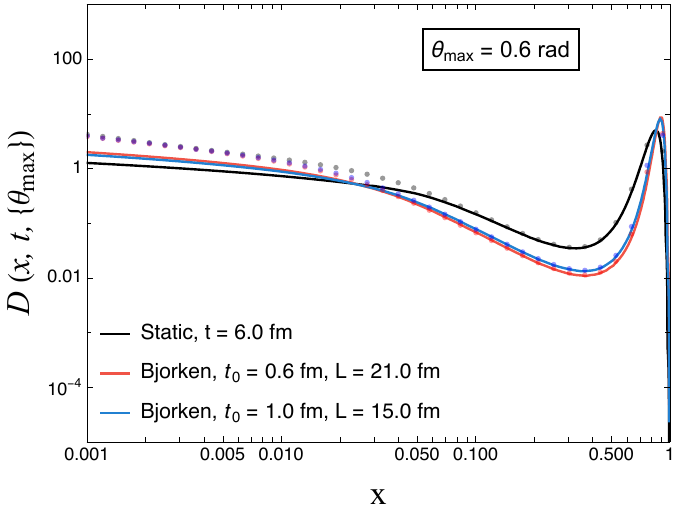}
\caption{The distributions $D(x,t;\{ \theta_{\rm max}\})$ for the $\theta_{\rm max}$ choices of $0.2$, $0.4$ and $0.6$ for effective medium length $L_{\rm eff} = 4\,$fm (left column) and $6\,$fm (right column). The faded dotted lines show the corresponding distribution for the full angular region and correspond to the curves shown in Fig.~\ref{fig:spectraDxt}.}\label{fig:Dvsxdifftheta}
\end{figure}

To understand the behavior of soft gluons in the low-$x$ regime, where the scaling estimates should work, we analyze the distribution of gluons in $x$ for a particular angular range. This corresponds to the angular integrated energy distribution 
\begin{equation}
    D(x,t; \{\theta_{\rm max} \}) = \int_0^{\theta_{\rm max}} \rmd\theta \,\bar{D}(x,\theta,t) \,.
\end{equation}
In Fig.~\ref{fig:Dvsxdifftheta}, we show the comparisons of the medium evolved spectra for different angular choices of $\theta_{\rm max} = 0.2,\, 0.4$ and $0.6$.  It is worth pointing out that the fully integrated over $\theta$ spectrum (dotted lines) was analyzed in the literature for purely radiative cascades in the cases of infinite media \cite{Jeon:2003gi,Blaizot:2013hx,Blaizot:2015jea,Mehtar-Tani:2018zba}, finite media \cite{Adhya:2019qse, Adhya:2021kws, Isaksen:2022pkj} as well as including in-medium collisional elastic scattering \cite{Kutak:2018dim,Iancu:2015uja,Schlichting:2020lef}. It was realized that a turbulent cascade, transferring energy from hard to soft modes, was responsible for the characteristic behavior in the low-$x$ regime.

In Fig.~\ref{fig:Dvsxdifftheta}, one can see the turbulent behavior in the region $0.3\le x\le 1$ for gluons within an opening angle $\theta_{\rm max}$. Since the hard part of the spectra ($x\sim 1$) is mainly driven by collinear splittings, we observe that the spectra remain the same with increasing the opening angle to a much broader cone of $\theta_{\rm max} = 0.4$ and $0.6$. Thus, the hard and the intermediate energetic gluons are largely confined to a narrow cone size. On increasing the medium size to $\Leff = 6\,$fm (right panels), we observe the depletion of the peak at high $x$, resulting in the transfer of more softer particles to the intermediate and small $x$ for all the angles. This is true for both the static and expanding finite media. The finite-size effect in the medium-induced emissions shows up around $x\sim 0.3$ (a dip) which would otherwise be much flatter for the infinite media. Next, in the softer region $(0.001\le x\le 0.3)$, the energy distribution caused by the accumulation of soft gluons towards the medium scale leads to a significant broadening effect on increasing the angular region from the narrow one of $\theta_{\rm max} = 0.2$ to the wider regions of $\theta_{\rm max} = 0.4$ and $0.6$. This region is driven mainly by radiative processes. Let us note that opening up the polar angle as well as increasing the medium size recovers more softer gluons for all the medium profiles. Interestingly, we observe the universality of the scaling between different medium profiles for all the jet opening angles as compared to the total angular integrated ones. In the following, we further probe the region of the soft gluons in angular space.

\begin{figure}[b!]
\centering     
\includegraphics[width=70mm]
{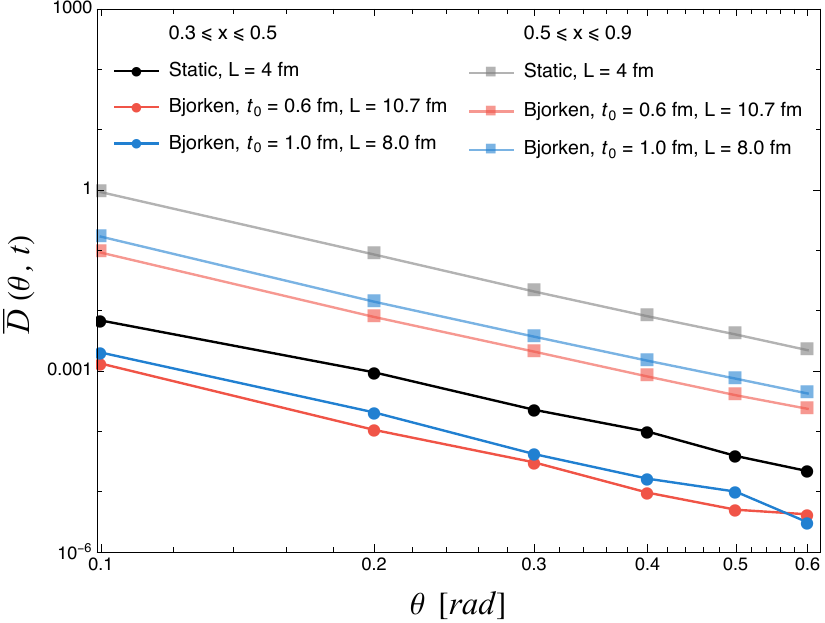}~~~~\includegraphics[width=70mm]{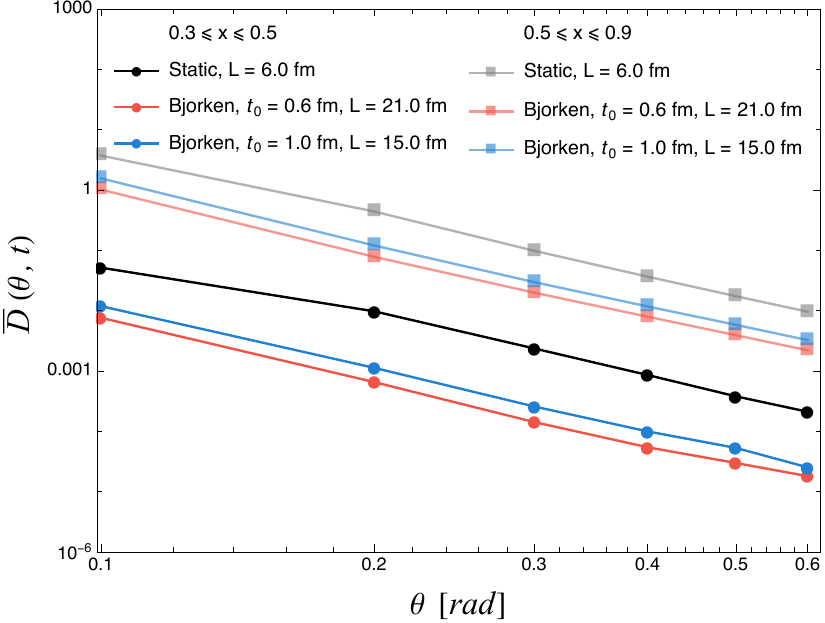}\\
\includegraphics[width=70mm]
{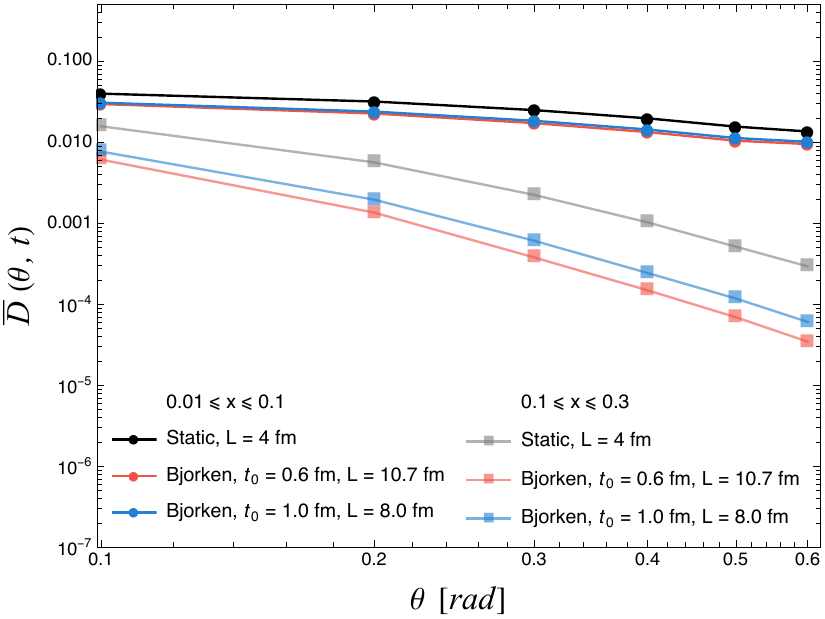}~~~~\includegraphics[width=70mm]{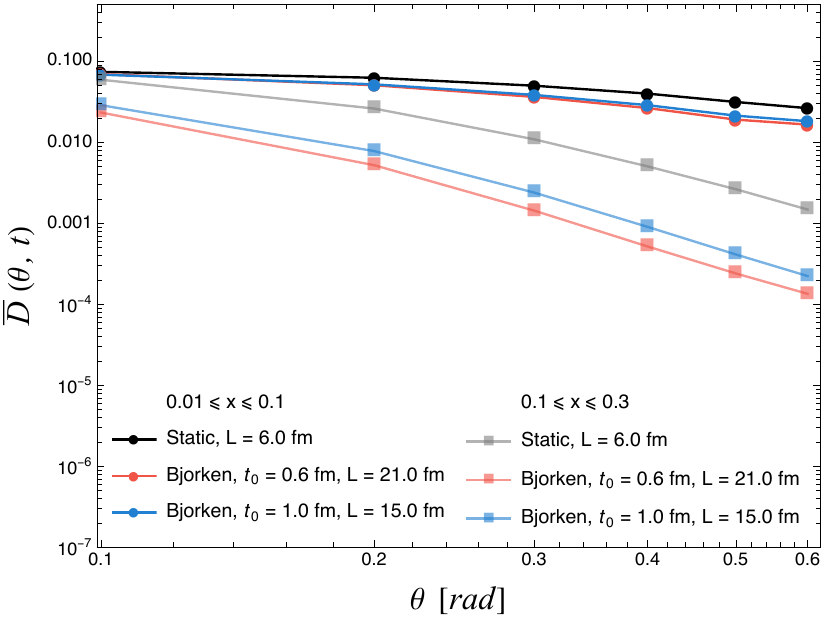}
\caption{The angular distributions $\bar{D}(\theta,t)$ in different ranges of $x$ for the effective medium length $L_{\rm eff} = 4\,$fm (left column) and $6\,$fm (right column). The upper panels show the distributions for the semi-hard and hard gluons, while the lower panels show the ones for the soft and semi-soft gluons.}
\label{fig:Dvsthetadiffx}
\end{figure}

Next, we analyze the angular dependence of the spectra for all the media profiles in four different $x$ ranges, restricting us again to the same equivalent in-medium path $\Leff$ for different medium scenarios. These are: $0.5 \le x \le 0.9$ (hard) and $0.3 \le x \le 0.5$ (semi-hard), which recover the gluons undergoing less in-medium quenching, $0.1 \le x \le 0.3$ (semi-soft) and $0.01 \le x \le 0.1$ (soft). For our purposes, we have chosen the angular region from $0.1$ to $0.6$ in $\theta$ to recover as much of the phase space of interest as possible. More importantly, this overlaps with with experimental studies of jet-quenching observables. Each panel in Fig.~\ref{fig:Dvsthetadiffx} shows the integrated energy distribution defined by
\begin{equation}
\label{eq:Dxtheta_xint}
    \bar{D}(\theta,t;\{x_{\rm min}, x_{\rm max}\}) = \int_{x_{\rm min}}^{x_{\rm max}} \rmd x\,\bar{D}(x,\theta,t) \,.
\end{equation}

In Fig.~\ref{fig:Dvsthetadiffx}, for the hard and semi-hard regimes (upper panels) we observe that the angular distribution drops sharply for hard gluons as the energy-momentum conservation imposes a restriction on the available phase space for large angle scatterings and tends towards the expected Coulomb tail $\sim 1/\theta^4$ \cite{Mehtar-Tani:2022zwf}. In all the panels, we observe the transverse momentum broadening effect where the energy is re-distributed to larger angles due to elastic scatterings with the medium quasi-particles as well as subsequent gluon splittings. In a larger length of the medium (right column), we observe a greater magnitude of the broadening as the jet spent more time within the media. 
More strikingly, for the hard, semi-hard, and semi-soft regimes included, the broadening effect is significantly larger in the static medium than in the expanding scenarios. Only in the genuinely soft regime do we observe a similar amount of broadening. 

The qualitative features of the results presented in Fig.~\ref{fig:Dvsthetadiffx} are as follows:
\begin{itemize}
    \item Collinear radiation causes depletion of the particles around the hard momentum sector $x \sim 0.5\,$--$\,0.9$. Subsequent splittings from hard particles to softer fragments undergo a successive broadening by which the energy is deposited in the soft sector of  $x \sim 0.1\,$--$\,0.01$. 
     \item The broadening effect is a combination of a decrease in momentum due to subsequent splittings as well as elastic collisions with the medium. The broadening effect is visualized as the energy deposited at small angles, $\theta = 0.2$, being transported to larger angles, $\theta = 0.4$ and $0.6$, out of the jet cone.
    \item We observe insignificant broadening of the
hard partons (a sub-leading effect compared to splittings), such that only the broadening near the medium scales $x \sim 0.01$ contributes to the energy at larges angles.
    \item The broadening effect in the soft sector is a nearly universal effect with respect to the static or expanding medium and the finite medium-size corrections. It is also universal with respect to the starting time for quenching of the Bjorken profile.
\end{itemize}

In order to estimate the magnitude of the transverse-momentum broadening effect, in the next section, we estimate more precisely the amount of the in-cone gluon fraction in the soft as well as the hard limit of the gluon momentum fraction.

\section{Soft versus hard gluons in a jet cone}
\label{sec:gluons-in-cone}

\begin{figure}[t!]
\centering     
\includegraphics[width=70mm]{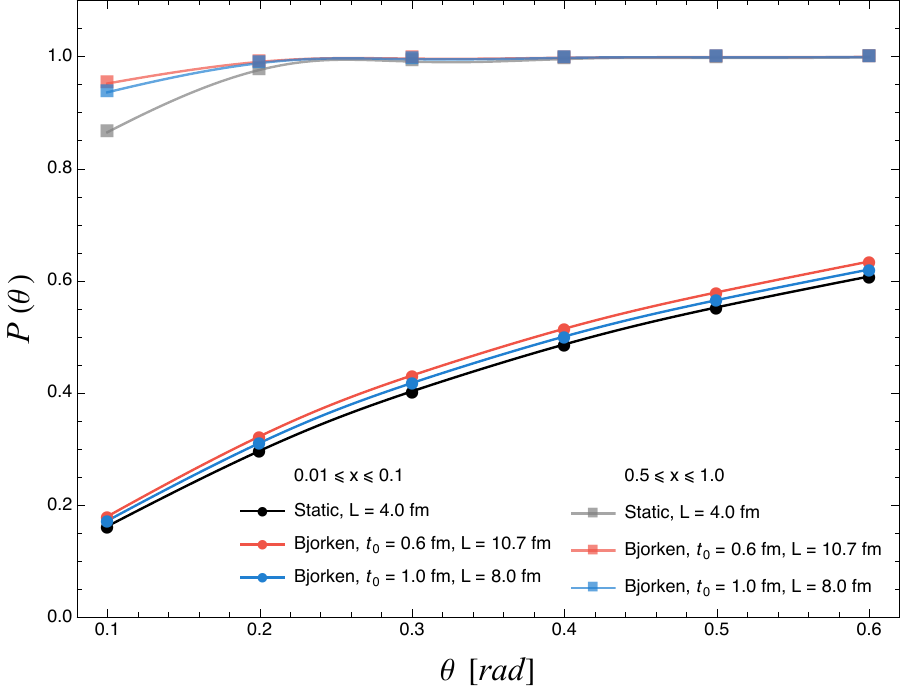}~~~~\includegraphics[width=70mm]{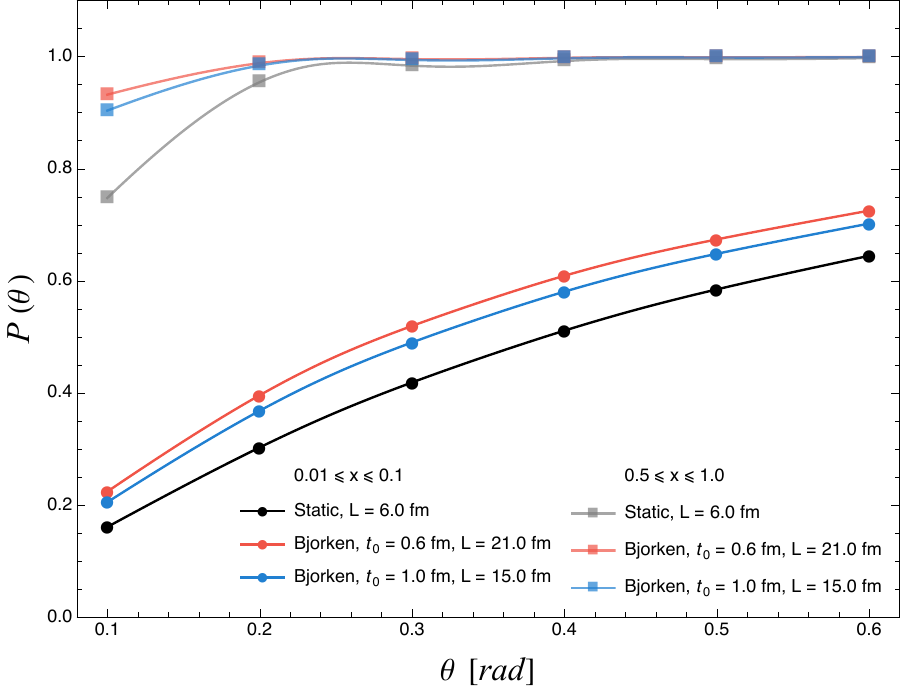}\\
\caption{Comparisons of the fraction of the gluon energy in a cone $P(\theta,t;\{x_{\rm min},x_{\rm max}\})$ as a function of its opening angle $\theta$ within different $x$-ranges, $x_{\rm min}\le x \le x_{\rm max}$, for the effective medium length $L_{\rm eff} = 4\,$fm (left panel) and $6\,$fm (right panel).
}
\label{fig:gfracxmax0p1}
\end{figure}
In this section, we study the features of the jet in-cone energy loss for the opening angle $\theta$ and as a function of the momentum fraction $x$. As before, we restrict ourselves to considering the same $\Leff$ in all three medium scenarios. In order to more precisely pin down the effect of medium expansion, let us define the fraction of the gluon energy contained inside a cone of the opening angle $\theta$ within the $x$-range $x_{\rm min}\le x \le x_{\rm max}$. This can be expressed as
\begin{equation}
P(\theta,t;\{x_{\rm min},x_{\rm max}\}) = \frac{\int_{x_{\rm min}}^{x_{\rm max}} \rmd x\int_{0}^{\theta} \rmd \theta'\, \bar{D}(x,\theta',t)}{\int_{x_{\rm min}}^{x_{\rm max}} \rmd x\int_{0}^{\pi} \rmd \theta'\, \bar{D}(x,\theta',t)}\,.
\end{equation} 
In the above equation, the denominator counts the total amount of energy contained in the chosen $x$-range for any angle. 
The numerical results are shown in Fig.~\ref{fig:gfracxmax0p1}, where we analyze the angular structure of the energy distribution of the gluons by dividing the medium evolved spectra into two regions corresponding to the large energy fraction: $ 0.5 \le x \le 1.0$ (square markers) and the soft sector: $ 0.01 \le x \le 0.1$ (round markers). 

For the hard sector (square markers), the fraction of energy in expanding media is already quite close to unity for very small cone angles. Nevertheless, the static medium recovers most of the energy already at $\theta =0.2$. Hence, for phenomenological studies, one does not expect to be very sensitive to the details of medium expansion.

In the soft sector (round markers), the gluon cascade has developed a significant width and one needs to go to large angles $\theta \gg 0.6$ to recover the full energy content. However, for $\Leff= 6\,$fm we clearly observe that the cascade is narrower in the expanding medium than in the static one (the situation for $\Leff = 4\,$fm is less clear). The potential to be sensitive to the details of the medium expansion through the width of the angular distribution at small-$x$ is interesting from a phenomenological point of view.

\section{Conclusions and outlook}
\label{sec:conclusions}

In this work, we have obtained spectra of jet particles that undergo scatterings and induced coherent splittings in a time-dependent medium, which is parameterized by the time-dependent transport coefficient $\hat{q}$ and the corresponding splitting kernels~\cite{Adhya:2019qse} of Eqs.~(\ref{eq:rate_static}) and~(\ref{eq:rate_bjorken}), corresponding to the static and Bjorken expanding medium, respectively.
Numerical results have been obtained using the {\sf MINCAS} Markov Chain Monte Carlo algorithm which calculates for the first time the corresponding fragmentation functions due to evolution via scatterings as well as medium-induced coherent splittings in the expanding medium.

Our findings can be summarized as follows:
\begin{itemize}
     \item The scaling approximations extracted from the singular limits of the purely radiative cascade work reasonably well for the medium-evolved spectra.
    \item We find that during the evolution of the energetic parton inside the media, hard partons remain effectively collinear and the momentum broadening is mainly caused by consecutive splittings rather than medium collisions and transverse-momentum exchanges. This type of behavior for hard partons is visible in the distributions of $x$ and $k_T$ in Fig.~\ref{fig:DxkT3D}, which follow at large $x$ and small $k_T$ largely the Gaussian distribution (due to the predominance of multiple soft scatterings), while at large $x$ and $k_T$, rare hard medium interactions have a significant influence, leading to a power-law behavior.
    \item Subsequent decrease of the momentum of these energetic partons into the soft momentum sector causes momentum broadening by transverse-momentum exchanges with the medium through elastic scattering as well as subsequent splittings into softer fragments. However, the broadening due to subsequent gluon splittings in the softer sector contributes to the out-of-cone energy loss at larger angles. This kind of behavior becomes apparent from Fig.~\ref{fig:Dvsxdifftheta}, where the differences between the full fragmentation functions $D(x,t)$ and its respective contributions within jet cones of different sizes are shown in comparison. It appears that with increasing the jet-cone angle, contributions of more and more soft particles are considered, while the behavior at large $x$ has almost no contributions due to large angles.
    \item The scaling approximation minimizes the difference for the soft jet fragments (small $x$) and for the same value of $L_{\rm eff}$ for the medium profiles as evident in Fig.~\ref{fig:Dvsxdifftheta}. The harder (large $x$) jet fragments differ among each other for different medium profiles for the same scaling approximation.
    \item The fraction of energy within a $x$-bin is also analyzed to distinguish between different medium profiles. In Fig.~\ref{fig:gfracxmax0p1}, we observe persistent differences in the soft sector, implying that the cascades in the expanding media are still relatively more collimated than their counterparts in the static scenarios.
    \item In the quantities we have chosen to analyze, we have not observed any sensitivity to the hydro initialization time $t_0$, in contrast to $v_2$ of jets \cite{Adhya:2021kws}, see also Ref.~\cite{Andres:2022bql}.
\end{itemize}

Naturally, for a similar in-medium path length in the static and expanding media, a leading parton will evolve much less due to the rapid diluting of medium density. This will lead to much fewer low-$x$ gluons \textsl{and} a narrower profile in the polar angle $\theta$. Comparing the distributions at the equivalent \textsl{effective} path length $\Leff$ should reduce this ``trivial'' effect. Nevertheless, our results indicate subtle differences between the developing profiles of the cascade in the static and expanding media in both the large- and small-$x$ regimes.

One of the natural extensions of the results presented in this paper is introducing the transverse-momentum-dependent splitting kernels for the individual medium profiles to solve the evolution equations for both the quark and gluon-initiated jets. 
Secondly, one should include the effects of thermal re-scattering whenever the energy of the fragments becomes comparable to the local medium temperature, i.e. $xp_0^+ \sim T$, see e.g.\ Ref.~\cite{Schlichting:2020lef}. Furthermore, one can study the role of these modifications on the radial distributions, or the so-called jet shape functions, of the jet-quenching by comparison with the recent data from the CMS \cite{CMS:2018zze} and ATLAS \cite{ATLAS:2019pid} experiments at the LHC. However, these studies are beyond the scope of the present paper and will be reported in a separate upcoming work.

\section*{Acknowledgements}
SPA and KK acknowledge the support of the Polish Academy of Sciences through the grant agreement PAN.BFD.S.BDN.612.022.2021-PASIFIC 1, QGPAnatomy. This work received funding from the European Union’s Horizon 2020 research and innovation program under the Maria Sk{\l}odowska-Curie grant agreement No.\ 847639 and from the Polish Ministry of Education and Science.
The research of MR was supported by the Polish National Science Centre (NCN) grant no.\ DEC-2021/05/X/ST2/01340.
KT is supported by a Starting Grant from Trond Mohn Foundation (BFS2018REK01) and the University of Bergen.

\bibliographystyle{jhep}
\bibliography{kTexpcades_v1}
\end{document}